# Enterprise Multi-Branch Database Synchronization with MSMQ

by

Emil Vassev

A Technical Article
April 2001

# Table of contents

# I. The Data Synchronization Problem

When we talk about databases there have always been problems concerning data synchronization. The latter is a technique for maintaining consistency among different copies of data (often called replicas). In general, there is no universal solution to this problem and often a particular situation requires a particular approach driven by specific conditions. This paper presents an approach tackling the issue of data synchronization in a distributed multi-branch enterprise database. The proposed solution is based on MSMQ (Microsoft Message Queue) [1, 2], a mechanism for asynchronous messaging. In this case study, we have about 200 local client-server stations termed as *branches*. On every branch, we have about 5 to 10 workstations and a server machine, where we can run our main application. Some of the branches are connected via LAN to the central station other are not. The database installed on each branch is local. We have also a central station with a *central database* that has to be synchronized with every *branch database*. All databases operate as independent self-contained units. Here, the transactions in branch databases do not depend on the transactions in other branch or central databases.

# II. Software Technologies

In this approach, we used the following software technologies:
- Windows 2000 [3] accompanied by the built-in services Microsoft Transaction Server (MTS or COM+) [1] and Microsoft Message Queue (MSMQ) [1, 2];
- SQL Server ver. 7.0 [4] as a database;
- Borland Delphi Enterprise ver. 5.0 [5] as a development tool.

# III. What is MSMQ?

MSMQ [1, 2] runs as a service [3] under Windows 2000, and exposes a set of objects that client applications and components can use to create messages and queues, and to send the messages to the appropriate queue. The MSMQ objects also provide methods for encrypting messages, acknowledging their receipt, and securing the queues.

An MSMQ enterprise is a set of computers that send and receive messages to and from a common set of queues. An enterprise can be divided into MSMQ sites. A site is typically a set of computers in the same physical location. MSMQ also lets you assign a set of computers to a connected network (CN). When a message is sent between two computers that do not belong to the same CN, the message must be routed through the computer configured to work as MSMQ Routing Server (MSMQ RS) [1, 2].

The first installation of MSMQ in any enterprise must be on a computer that will serve as the Primary Enterprise Controller [1, 2].

## IV. Using Transactional Messages with MSMQ

When you send a message inside a transaction, you get a few guarantees that you don't get with a non-transactional message. First, MSMQ provides exactly-once delivery semantics. It takes some extra precautions so that messages that are on their way to a destination queue are not lost or duplicated. MSMQ also ensures that messages inside the same transaction are delivered in the order in which they were sent.

In MSMQ, transactional messages must be sent to transactional queues. You can't change the transactional attribute after a queue has been created. You can send and receive messages with MSMQ in two ways. You can use MSMQ's internal transaction mechanism, or you can use external transactions that are coordinated by the Distributed Transaction Coordinator (DTC) [6]. In our case it will be MTS [1, 6] (COM+ for Windows 2000). The internal transaction mechanism of MSMQ provides best performance, however, when you use internal transactions, MSMQ can't coordinate the transaction with any other type of resource manager, such a SQL Server database. External transactions are known also as distributed transactions because they use DTC. They let you define transactions that include message passing along with operations to other resource managers (SQL Server). For example, you can write a transaction that receives a request message, modifies the database, and sends a response message.

## V. MSMQ and Database Synchronization

The core of this synchronization solution is the execution of the SQL statements (delete, update, and create SLQ query statements), which passed well on the branch servers, on the central server. This helps us have online, real time, branch-server synchronization when there is *online connection*, and forces low-latency synchronization for branches that are not currently connected (*offline connection*). In other words, the synchronization delay will depend on the connection type and speed (see Fig. 1).

The database synchronization in this solution uses the MSMQ message delivery technique, which assures exactly once-delivery semantics. Here, on each BRANCH and CENTRAL servers run special synchronization processes (SPB for BRANCH - Fig. 2, and SPC for CENTRAL – Fig. 3). Together, with MSMQ, they are responsible for message delivering. MSMQ dispatches messages between two remote machines (Branch and Central) when there is an established connection. By reason, the messages that arrived on CENTRAL or BRANCH are well passed SQL statements (committed SQL statements), when SPC or SPB execute them on CENTRAL and BRANCH we will have a data equality provided by this SQL statement. This process is two-direction process, and every message acceptor (CENTRAL or BRANCH) can be message sender as well. All the machines (PC stations) must have an IP address visible by the CENTRAL and a MSMQ independent client installation. When DB Synchronization works over the Internet, eventually, we need Virtual Private Network (VPN) [3] to ensure a static IP address range for the connected machines.

## VI. Synchronization Process on Branch Servers

The Branch Server Synchronization Mechanism is designed to work in the system's standard operating mode (log-on mode) as well as in log-off mode.

### 1. Process Components

The synchronization process on each BRANCH is distributed among the following components (see Fig. 2):
- **Synchronization NT Service (SNTS) application** – Windows NT Service [3] application, which is responsible to listen to the MSMQ incoming port for newly arrived messages from the Head Office (Remote Server – CENTRAL). This application takes the lifetime control over the COM+ [6, 7] components. SNTS has dependencies on the system services MSSQL Server and Message Queuing [1]. Hence, SNTS will not be able to run without these two services. Note that as an NT Service SNTS is capable of working when the system is in log-off mode.
- **Agent BRANCH** – monitoring tool that provides the users with information about the whole synchronization process and gives some tools for maintenance and support. Also, it is an element of the dispatching process – MTS triggers the dispatching process and gets all pending db records that have to be dispatched as messages to the CENTRAL. This application takes the lifetime control over the COM+ components within the Dispatching Process.
- **Two COM+ components** installed in COM+ [6, 7] library packages (they use the client address space). They are responsible for the sophisticated transactional process implementation: reading and writing to database, sending and receiving messages to and from MSMQ. The first one (see MTSMSMQController in Fig. 2) controls the transactions with DTC [6] and the second one (see MTSMSMQService in Fig. 2) proceeds with the actions.
- **MSMQ private queues and virtual outgoing queues** (their lifetime depends on the network connection – LAN, the Internet or RAS). When a message is sent to a remote queue MSMQ creates automatically an outgoing queue, which acts as a bridge between the remote machines. This ensures the messages will be delivered to the destination queue when there is established connection.

### 2. Process Flow

- **Receiving process** (Incoming Process CENTRAL – BRANCH):
  A newly arrived from CENTRAL message triggers an OnArrived event, which notifies SNTS. It calls the COM+ objects. The first one (MTSMSMQController) creates a DTC (Distributed Transaction Coordinator) transaction context [1, 6] and within this context creates an instance of MTSMSMQService. This COM+ object gets the message from the queue, parses the message body, saves it to db, and executes the SQL query (part of the message) if there is permission for that.

- **Dispatching process (Outgoing Process BRANCH – CENTRAL):**
  When the main application (see MainApp in Fig. 2) is successfully done with a db transaction it registers the executed SQL query(s) in a special table and sends a Windows API message [3] to the Agent BRANCH application. Actually, this message triggers the Dispatching process on the branch machine.

Agent BRANCH (AB) takes all pending records from the query table. For every record, AB activates the COM+ components [6], which send a message to the CENTRAL-specified MSMQ queue and update the status of that record.

## VII. Synchronization Process on Central Server

The Central Server Synchronization Mechanism is designed to work in the system's standard operating mode (log-on mode) as well as in log-off mode.

**1. Process Components**

The synchronization process on the CENTRAL is distributed among the following components (see Fig. 3):
- **Synchronization NT Service** (SNTS) application – Windows NT Service [3] application, which is responsible to listen to the MSMQ [1, 2] incoming ports for newly arrived messages from the branch servers (Remote Servers – BRANCH). This application takes the lifetime control over the COM+ components. SNTS has dependencies on the system services MSSQL Server and Message Queuing [1]. Hence, SNTS will be not able to work without these two services. As NT Service, SNTS is capable of working when the system is in log off mode.
- **Agent CENTRAL** – monitoring tool that provides the users with information about the whole synchronization process and gives some tools for maintenance and support. Also, it is an element of the dispatching process – MTS triggers the dispatching process and gets all pending db records, which have to be dispatched as messages to the BRANCHES. This application takes the lifetime control over the COM+ objects within the Dispatching Process.
- **Two COM+ components** installed within COM+ [6] library packages (they use the client address space). They are responsible for the sophisticated transactional process implementation: reading and writing to database, sending and receiving messages to and from MSMQ [1, 2]. The first one (see MTSMSMQController in Fig. 3) controls the transactions with DTC [6], and the second one (see MTSMSMQService in Fig. 3) proceeds with the actions.
- **MSMQ private queues and virtual outgoing queues** (their lifetime depends on the network connection – LAN, the Internet or RAS). When a message is sent to a remote queue, MSMQ automatically creates an outgoing queue, which acts as a bridge between the remote machines. This ensures the messages will be delivered to the destination queue when there is established connection.

**2. Process Flow**

- **Receiving process** (Incoming Process CENTRAL – BRANCH):
  A newly arrived from a BRANCH message triggers an OnArrived event, which notifies SNTS. It calls the COM+ [6] objects. The first one (see MTSMSMQController in Fig. 3) creates a DTC (Distributed Transaction Coordinator) [6] transaction context and within this context creates an instance of MTSMSMQService. This COM+ object gets the message from the queue, parses the message body, saves it to db, and executes the SQL query (part of the message) if there is permission for that.
- **Dispatching process** (Outgoing Process CENTRAL - BRANCH):
  When the main application (see MainApp in Fig. 3) is successfully done with a db transaction it registers the executed SQL query(s) in a special table and sends a windows API message [3] to the Agent CENTRAL application. Actually, this message triggers the Dispatching Process on the Central machine.

Agent CENTRAL (AC) takes all pending records from the query table. For every record, AC activates the COM+ objects, which send a message to a BRANCH-specified MSMQ queue and update the status of that record.

# VIII. MSMQ Mailing System

In this section, an enterprise mailing system is proposed based on MSMQ. The MSMQ mailing system (MMS) will serve as an e-mail service among all servers (branches and central). This system will be capable of transferring as well text messages as attached files regardless their format. Actually, that can be a secure corporative mail-system. The messages can be encrypted for safety reason by using the most advanced hashing and encrypting algorithms, which are available for internal use in MSMQ. Sending a message from a branch to another one will pass through the dispatching center on CENTRAL (see Fig. 4), which has connections with all machines in the Enterprise. The coop-work with the firewall MSMQ capabilities ensures access to queue's machines through a firewall. If the local system has an Intranet service, the newly received MSMQ mail can be dispatched as an e-mail to the every particular e-mail address. SMMS will be able to work in a cooperative mode with Internet/Intranet, but that does not mean that it is a necessary condition. By using special "journal queues" MMS can track all messages sent to and received from a user machine. This is similar to the most e-mail systems, where users can keep a copy of everything sent and received. For user convenience, MMS can be integrated within Word and Excel, where messages may be sent to recent recipients with a single command. The mailing system will be independent of the synchronization process, but it can be used for synchronization acknowledgement.

## References

[1] Alex Homer and David Sussman. Professional MTS and MSMQ Programming with VB and ASP. John Wiley & Sons, 1998

[2] Rhys Lewis. Advanced Messaging Applications with MSMQ and MQSeries. Que, 1999

[3] David A. Solomon and Mark E. Russinovich. Inside Microsoft Windows 2000, 3rd Edition. Microsoft Press, Redmond, Washington, USA, 2000

[4] Ron Soukup and Kalen Delaney. Inside Microsoft SQL Server 7.0. Microsoft Press, Redmond, Washington, USA, 1999

[5] Inprise Corporation. Borland Delphi 5 for Windows 98, Windows 95, & Windows NT – Developer's Guide. Inprise Corporation, 1999

[6] Don Box, Keith Brown, Tim Ewald, and Chris Sells. Effective COM: 50 Ways to Improve Your COM and MTS-based Applications. Addison-Wesley Professional, 1998

[7] Dale Rogerson. Inside Com - Microsoft Programming Series. Microsoft Press, 1997

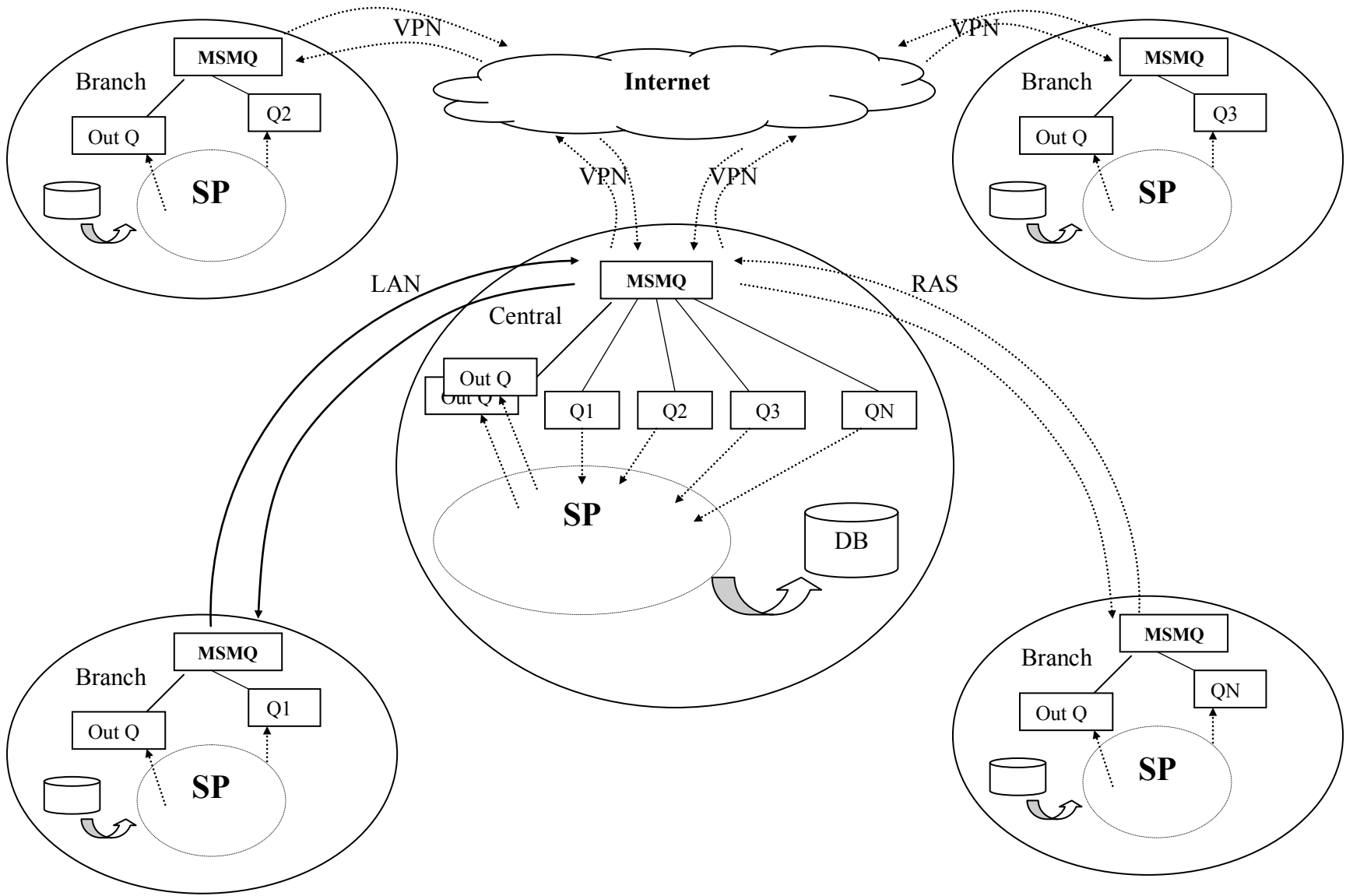

Fig. 1. Local and Central DB Synchronization

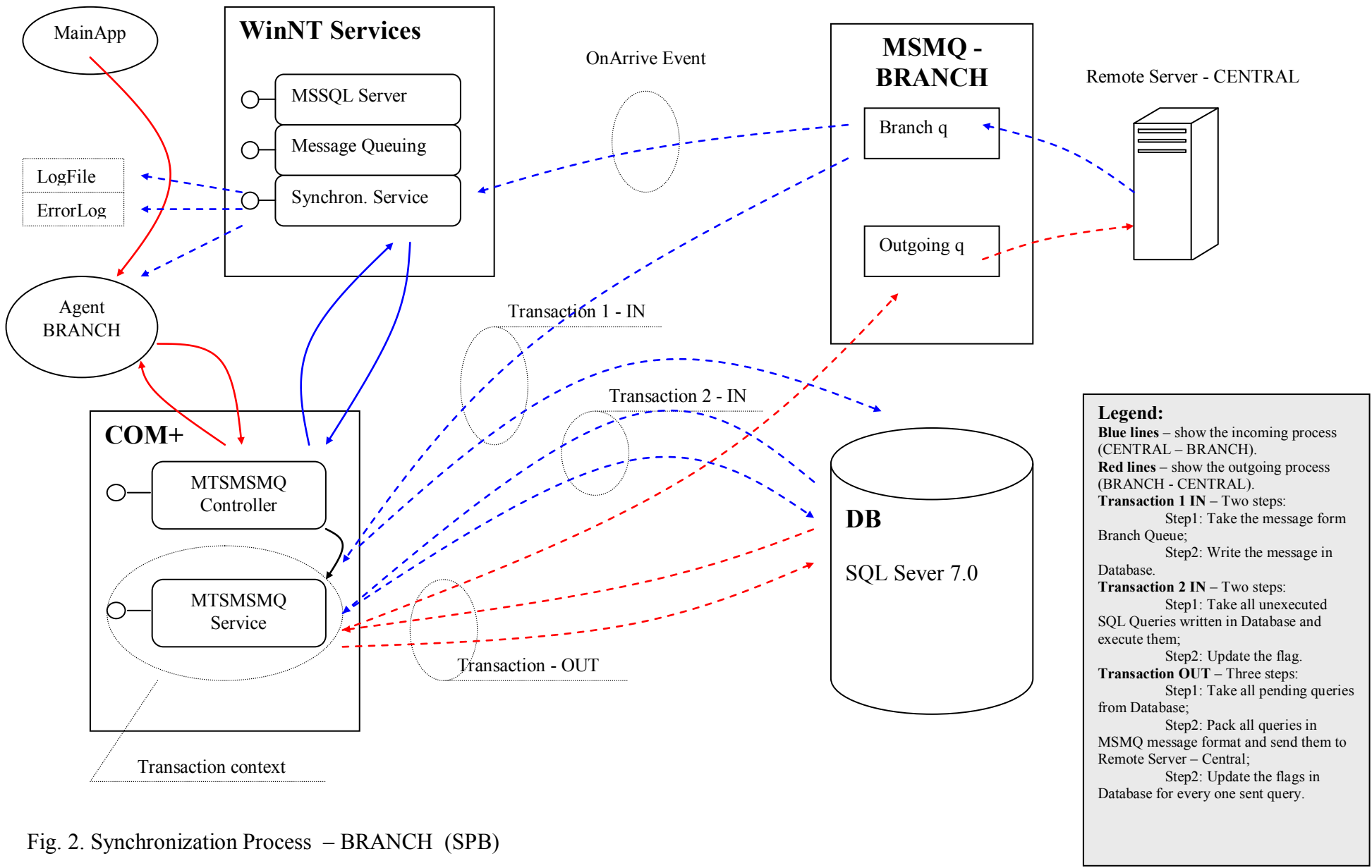

Fig. 2. Synchronization Process – BRANCH (SPB)

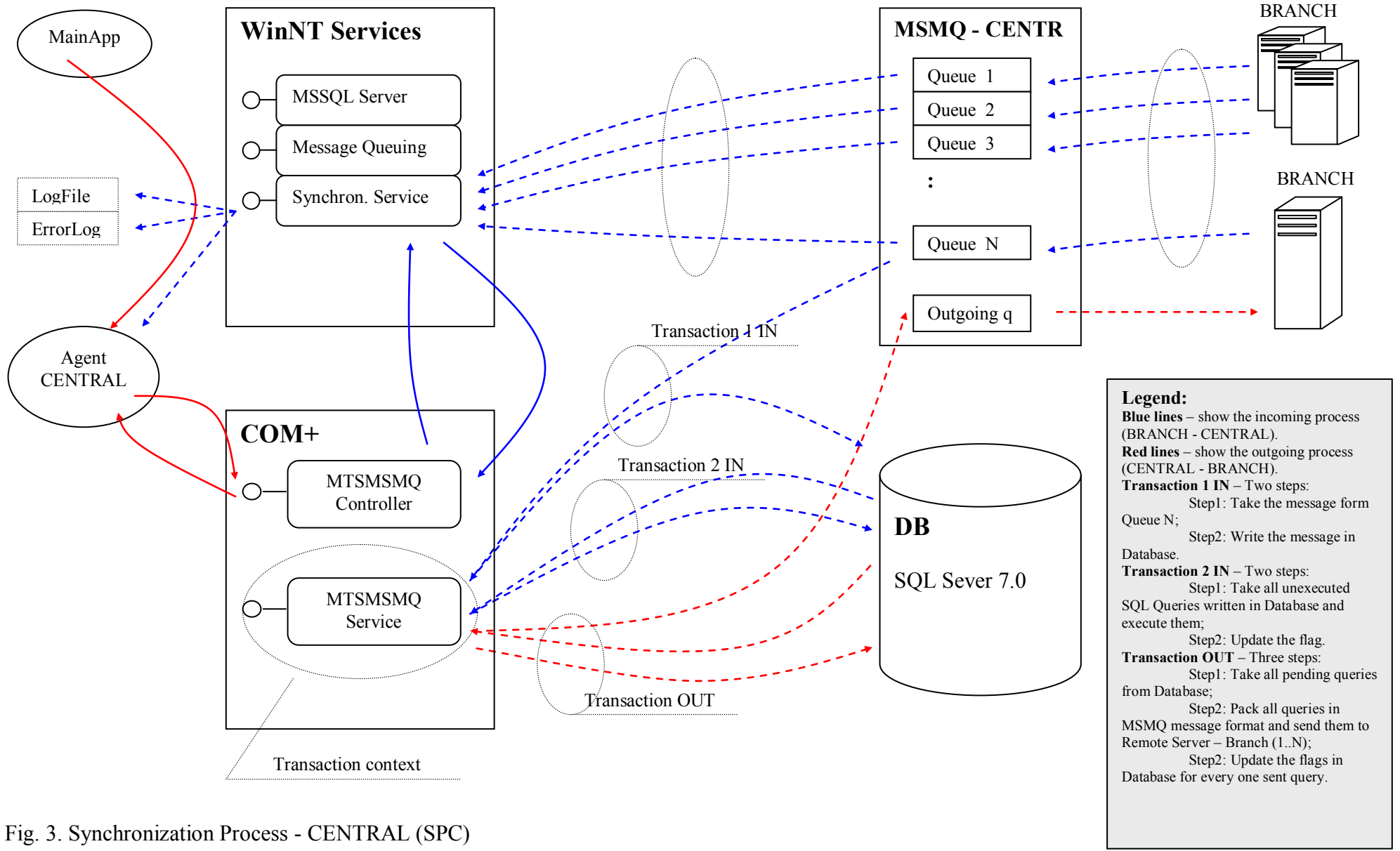

Fig. 3. Synchronization Process - CENTRAL (SPC)

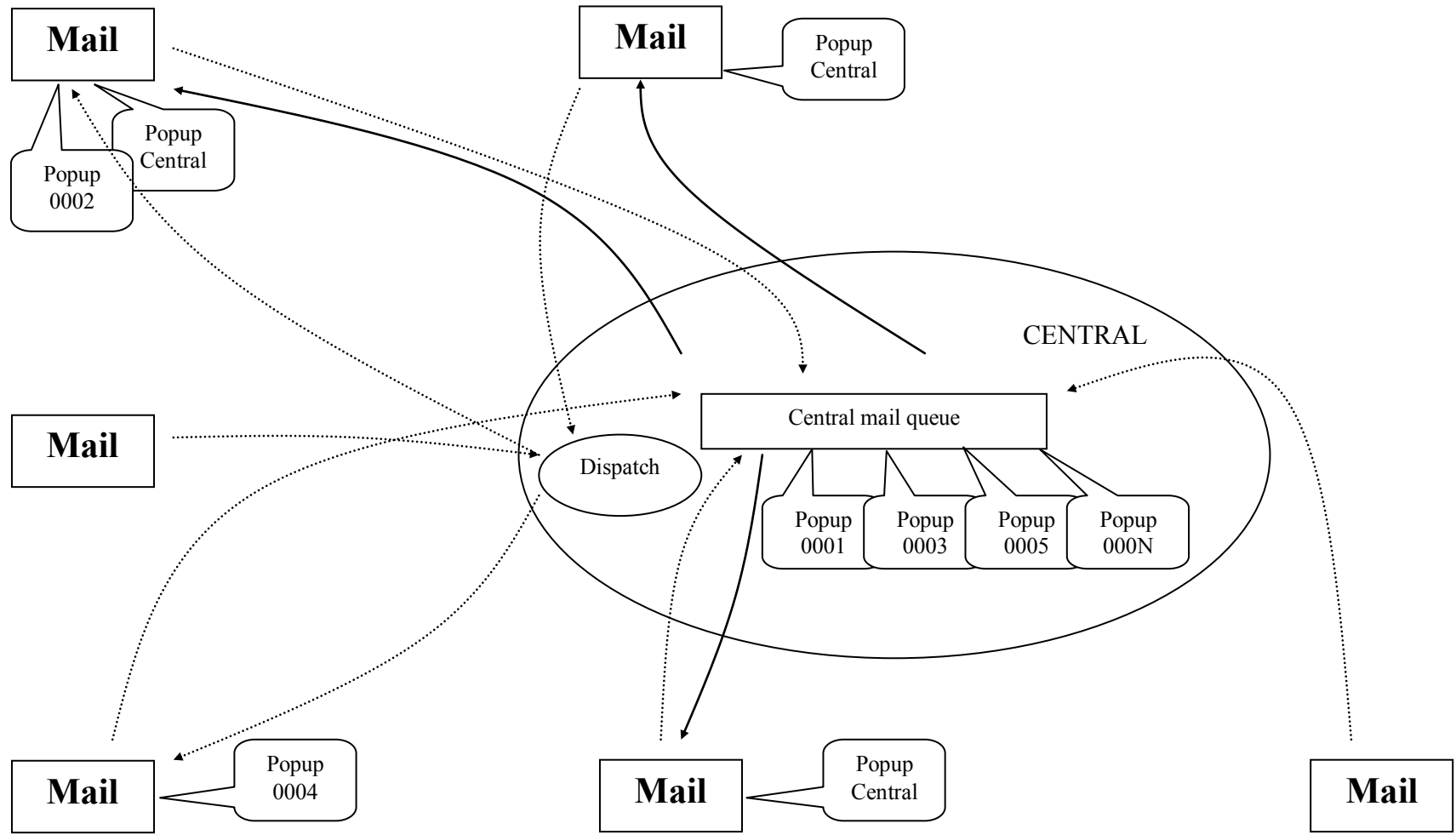

Fig. 4. MSMQ Mailing System